\documentclass[twocolumn,amsmath,amssymb,prd]{revtex4}

\usepackage{epsfig}
\usepackage{graphicx}
\usepackage{dcolumn}
\usepackage{latexsym}
\usepackage{amssymb}

\def \be{\begin{equation}}
\def \ee{\end{equation}}
\def \bea{\begin{eqnarray}}
\def \eea{\end{eqnarray}}

\begin{document}


\title{Hybrid Monte Carlo with Wilson Dirac operator on the Fermi GPU}

\author{Abhijit Chakrabarty}
\affiliation{Electra Design Automation, SDF Building, SaltLake Sec-V, Kolkata - 700091.}
\email{krishnanagar@gmail.com}
\author{Pushan Majumdar}
\affiliation{Dept. of Theoretical Physics, Indian Association for the Cultivation of 
Science, Jadavpur, Kolkata - 700032.}
\email{tppm@iacs.res.in}

\begin{abstract}
In this article we present our implementation of a Hybrid Monte Carlo algorithm for 
Lattice Gauge Theory using two degenerate flavours of Wilson-Dirac fermions on a 
Fermi GPU. We find that using registers instead of global memory speeds up the code 
by almost an order of magnitude. To map the array variables to scalars, so that the 
compiler puts them in the registers, we use code generators. Our final program is 
more than 10 times faster than a generic single CPU.
\end{abstract}
\maketitle


\section{Introduction}
Lattice gauge theory \cite{schol} is a formulation of gauge theory on a 
Euclidean space-time lattice $\Lambda$. $\Lambda$ consists of $N_{\rm site}$ 
points $n$ ordered in some fashion and $d\times N_{\rm site}$ links 
$(n,\mu)$ where $d$ is the number of space-time dimensions.
This formulation of gauge theory is amenable to computer simulations.
Lattice gauge theory simulations can perform non-perturbative computations
 for static observables in Quantum 
Chromodynamics (QCD). The expectation value of an observable ${\mathcal O}$ 
can be obtained by evaluating a Euclidean path integral as
\be
\langle {\mathcal O}\rangle = \int [DU]d\psi_i d\bar\psi_i \,{\mathcal O}\, 
e^{-\bar\psi_i\left ( D^{\,(U)}+ m_i\right )\psi_i -S_g(U)}.
\ee
$U_{n,\mu} \in SU(N_c)$ is a group valued matrix \footnote{In QCD $N_c=3$
but other values of $N_c$ are also theoretically interesting.} defined on the
link $(n,\mu)$ and plays the role of a parallel transporter (gauge field).
$\psi_i$ represents a fermionic species with mass $m_i$
and is defined at each lattice point $n$.
$D$ is a lattice Dirac operator and $S_g$ 
is a discretized version of the continuum Yang-Mills lagrangian 
$\frac{1}{2}{\rm tr}F_{\mu\nu}F^{\mu\nu}$. 

We first integrate out the fermionic fields to obtain 
\be
\langle {\mathcal O}\rangle = \int [DU] \prod_i \det \left ( D^{\,(U)}+ m_i\right ) 
\,{\mathcal O}\,e^{-S_g(U)}.
\ee
Computing the determinant is an $O(N^3)$ process for a $N\times N$ matrix and for the 
4-dimensional Dirac oprator $N=N_c\times 4 \times N_{\rm site}$. Even for moderate 
lattice sizes $N\sim 10^6$. An explicit 
evaluation of this determinant is not feasible. It is usually 
re-exponentiated by introducing additional bosonic fields $\phi$ called pseudofermions as 
\be
\langle {\mathcal O}\rangle = \int [DU]d\phi^{\dagger}_i d\phi_i \,{\mathcal O}\,
e^{-\phi^{\dagger}_i\left ( D^{\,(U)} + m_i\right )^{-1}\phi_i ~-S_g(U)}.
\ee
The path integral now involves evaluating the inverse of the matrix 
$\left ( D^{\,(U)}+ m_i\right )$ on the vector $\phi_i$. This can be done in at most 
$O(N)$ steps using conjugate gradient (CG) type of algorithms.

Though lattice gauge theory simulations are computationally expensive, they can be easily parallelized. 
Such simulations can therefore be efficiently carried out on a Graphics Processing 
Units (GPUs) which have about 500 compute cores. Since the first implementation of lattice gauge 
simulations on GPUs 
in 2007 \cite{qcdvideo} there has been numerous implementations of various formulations of 
lattice gauge theory on GPUs. A collection of GPU subroutines for several lattice actions has been 
put together in a package called QUDA \cite{quda}. Implementation of pure gauge codes and spin 
models on GPUs have been reported in \cite{puregauge} and \cite{spin}. 

In this article we explore a new method of implementing lattice simulations on the GPU 
using register variables and avoid the slow global memory for storing run-time variables. 
We consider one of the most popular formulations namely lattice gauge
theory with two mass degenerate fermions. 
This partition function can be written as 
\be
{\cal Z}=\int [DU]d\phi^{\dagger} d\phi \,e^{-\phi^{\dagger}\!\left (\frac{1}{{\cal M}^{\dagger}
(U){\cal M}(U)}\right ) \phi ~-S_g(U)}.
\ee
where ${\cal M}(U)=D^{\,(U)}+ m$

In section 2 we briefly describe the algorithm which we use. Section 3 describes the 
architecture of the GPU cards on which the program was tested and discusses the 
modifications we have to make in the usual CPU program for efficient running on the GPU. Finally 
in section 4 we report on the performance of the program and our conclusions.  

\section{The HMC algorithm}

Currently the most efficient algorithm for generating gauge configurations with 
dynamical quarks on the lattice is known as Hybrid Monte Carlo (HMC) \cite{hmc}.
In this algorithm the dynamics
is generated by equations of motion corresponding to the Hamiltonian
\be
{\cal H}(U,p)=\frac{1}{2}\,{\rm tr}\,p^2 +S_g(U)+\phi^{\dag}\left (
\frac{1}{{\cal M}^{\dagger}(U){\cal M}(U)}\right )\phi.
\ee
Here $p_{n,\mu}\in su(N_c)$ is an auxiliary momentum variable with $U$ and 
${\rm e}^{ip}$ being conjugate to each other. $p$ is drawn from a Gaussian distribution 
and $\phi={\cal M}^{\dagger}(U)\chi$ where $\chi$ is also 
drawn from a Gaussian distribution. Both $p$ and $\chi$ are periodically refreshed.

The equations of motion (molecular dynamics equations) define a trajectory in the 
phase space. Integrating these equations numerically
is the most time consuming part in the algorithm accounting for 80 to 90\% of the computation
time depending on simulation parameters. This part of the algorithm
involves repeated multiplication of a matrix ${\cal M}=\left ( D + m\right )$
on the vector $\phi$. Parallelization of the matrix vector multiplication speeds up
lattice gauge theory simulations to a great extent.

The HMC algorithm thus refreshes the momenta periodically to ensure ergodicity,
uses the Hamiltonian equations of motion to move quickly
through the phase space and perofrms a Monte Carlo accept/reject step to correct
for the finite time step used in the integration. It is an exact algorithm.

\section{Fermi architecture and organization of the GPU program}

Our program was tested on two GPU cards the C2050 and X2090 (different versions of 
the NVIDIA Fermi Card) and we briefly discuss their architecture below.
Both cards have intrinsic double precision support and are CUDA compute version 2. 

The computation on the cards are carried out by multiprocessors each of which have 
32 cores and a maximum of 64 KBytes of local storage (shared memory + L1 cache).
The C2050 has 14 such multiprocessors giving it a total of 448 cores while the X2090 
has 16 such multiprocessors giving it a total of 512 cores. 

Nvidia GPU cards have several different categories of memory and their access times vary widely.
The C2050 has a global memory of 3 GB while the X2090 has a global memory of 6 GB. These are the 
largest memories but 
their access times are between 100 and 150 clock cycles. Every multiprocessor 
on the other hand has 32768 4-byte registers whose combined memory capacity is 128 Kbytes and 
access to the register variables takes only 1 clock cycle. Each thread can have a maximum of 
63 registers. Fermi cards have the added feature that registers spill over to the 
shared memory which too can be accessed in 2 - 3 clock cycles. The combined storage capacity 
of registers and shared memory is about 192 Kbytes per multiprocessor or about 2.7 MB 
for the C2050 and 3 MB for the X2090. In our program we have attempted to use the registers 
as much as possible instead of the global memory.      

The CUDA compiler usually tries to place all scalar variables in the register \cite{cudaman} 
while arrays are typically placed in the global memory. 
In HMC, the molecular dynamics part deals mainly with three large arrays: the force which is an
array of dimension $N_c,N_c,4,N_{\rm site}$ and the momenta and the pseudo-fermion fields, both
of which are arrays of dimension $N_c,4,N_{\rm site}.$
Our main challenge therefore is to make sure that the elements of these arrays are 
placed in the register and not in the global memory.

The structure of the program is roughly the same for the CPU and the GPU.
There is an initialization step where the gauge field is initialized.
The gauge field is then evolved through a certain number of trajectories. At the 
end of each trajectory measurements are carried out on the gauge configuration. 

The evolution of the gauge fields consists of the following steps for each 
trajectory.
\begin{enumerate}
\item Setting up the pseudofermion fields and momenta.
\item Integrating the molecular dynamics equations of motion by the leap-frog method. 
This sets up the proposed configuration.
\item  Accepting or Rejecting the proposed configuration depending on ${\cal H}_{\rm new} 
- {\cal H}_{\rm old}$, where ${\cal H}_{\rm old}$ is the Hamiltonian at the beginning of the 
trajectory and ${\cal H}_{\rm new}$ is the Hamiltonian at the end of the trajectory. 
\end{enumerate}

Communication between the CPU and the GPU is through a PCI bus. A major bottleneck if 
one frequently transfers data between them. To keep a check on this overhead, we tried to 
strike a balance between the time taken for transferring the data and the computational time on the GPU.
Since all the input and output is carried out 
by the CPU we had to transfer the data back to the CPU at the end of each trajectory.

For the computation intensive parts we wrote CUDA kernels (enumerated below) which carried out the 
following computations in parallel on the GPU.
\begin{enumerate}
\item The result of operating the Dirac operator and its hermitian conjugate on 
the pseudofermion field $-$ a matrix vector multiplication on a vector of size 
$N_c\times 4 \times N_{\rm site}$.
\item The driving force for the molecular dynamics equation due to the gauge fields.
\item The driving force for the molecular dynamics equation due to the fermion fields 
and evolving the momenta.
\item Evolution of the gauge fields.
\end{enumerate}

We now briefly describe the essential modifications to the CPU subroutines for efficient running of the 
corresponding GPU kernels.

We first try to optimize the two matrix-vector multiplication routines which 
compute the action of the Dirac operator and its hermitian conjugate on the pseudofermion field as a part
of the conjugate gradient routine. So we 
first try to optimize this routine. The Dirac operator requires four constant $4\times 4$ matrices 
($\gamma_{\mu}$) for 
its definition. Only 8 elements of these matrices are non-zero. We explicitly
work out the result of the matrix-vector multiplication for the $\gamma$ matrices retaining only the 
non-zero elements. Thus the loop over the dirac indices is fully unrolled.
 This however is not special to the CUDA kernels and is done also for the CPU subroutines.

Generally we use the available CUDA BLAS functions as much as possible. However for the action of the Dirac 
operator on the pseudofermion field, it is not possible to use the BLAS matrix-vector multiplication routines 
as the matrix representing the Dirac operator is too large to be stored in either the CPU or the 
GPU memory. So we store the source and the resultant vectors and compute the necessary matrix elements on the fly.  
For this kernel, the main optimization task was to break down the array representing the pseudofermion 
field into scalar variables. Since the pseudofermion field typically consists of $\sim 10^6$ elements, it is 
 impossible to do this by hand. We therefore wrote a code generator which runs over the necessary 
loops and yields the equations for the resultant vector in terms of independent 
scalar variables. A naming convention we adopt is to name the scalar variables by concatenating 
the array indices with the vector name. 

The second computationally intensive task is computing the driving force for the molecular dynamics 
equation due to the fermion fields. On a single thread, they typically consume 15\% of the trajectory time. 
After parallelization of the matrix-vector multiplication routines on the GPU, this was consuming the 
largest amount of 
time. To optimize this routine, we followed the same strategy as in the matrix-vector multiplication case. 
Unrolling the dirac loops in the force calculation is slightly more involved as they involve factors like 
$\gamma_{\mu}\gamma_{\nu}$. However a bit of algebra again allows us to retain only the non-zero 
elements and with the help of a code generator we automatically wrote down the momenta evolution equations. 
On the CPU this routine consumed significantly less time compared to the conjugate gradient routine. So this 
optimization was not carried out on the corresponding CPU subroutines.   

In addition to these optimizations, we also had a routine to compute the neighbour indices 
on the fly to avoid storing the neighbour array on the GPU.

The program is available from the authors on request.

\section{Performance and Discussions}
The speed-up of the GPU program was primarily judged by its performance on a C2050 Nvidia GPU card 
running at 1.15 GHz vis a vis 
its performance on an Opteron cluster with QDR Infiniband interconnect.  
Each node of the cluster consisted of two 6-core 2.2 GHz CPUs. We also have 
comparative results for one lattice size, viz. $24^4$, on a CRAY XE6 and a X2090 Nvidia GPU 
card running at 1.3 GHz on a CRAY XK6 node.

On the Opteron cluster the highest performance was obtained for one MPI process 
per node. We therefore fixed the number of MPI processes to 8 (number of available nodes) and varied the number 
of OpenMP threads in each node. For the sake of uniformity, the same test run pattern was followed even on the 
Cray.
In addition to the total time taken, 
timings for each conjugate gradient call and fermionic force calculations (the two most time consuming 
subroutines) were written out.

\begin{table}
\begin{tabular}{|c|c|c|c|c|}
\hline
Lattice& GPU & OMP & SINGLE & MPI+OMP \\
\hline
8$^4$ & 2691.084    &  4264.125   & 12551.820 &  3725.766 \\
16$^4$ & 37555.429  & 84715.792   & 409478.538 & 74381.869 \\
24$^4$ & 368173.325  & 999445.502  & 3717039.929 & 376495.740 \\
28$^4$ & 698531.978   & 1567117.214  & 7000000.000$^*$ & 1242310.046 \\
\hline
\end{tabular}
\caption{Timing (in seconds) of the program on a C2050 GPU, single CPU, SMP (12 threads) and 
mixed MPI and OMP with 8 MPI processes each having 12 threads for different lattice volumes.
These runs were performed on an Opteron cluster with QDR Infiniband interconnect.
($^*$estimated)}\label{time}
\end{table}

\begin{table}
\begin{tabular}{|c|r|r|c|c|}
\hline
Lattice & GPU & OMP & SINGLE & MPI+OMP \\
\hline
8$^4$ & 4.66   &  2.94  & 1 & 3.370 \\
16$^4$ & 10.90 & 4.83   & 1 & 5.505 \\
24$^4$ & 10.10  & 3.72  & 1 & 9.873 \\
28$^4$ & 10.02  & 4.47  & 1 & 5.630 \\
\hline
\end{tabular}
\caption{Same as table \ref{time}, but showing the relative speed-up.}\label{speedup}
\end{table}

In table \ref{time} we record the time taken for the various lattice volumes using different 
compute modes of the Opteron cluster and the C2050 GPU card mounted in the Opteron cluster.
These volume scaling runs were performed at $\beta=5.5$ and $\kappa=0.48$ with a time step of 0.2 and 
50 integrations per trajectory. The runs were for 200 trajectories each.
The timing for the single CPU run for the $28^4$ lattice is an estimate based on the $V^{5/4}$ scaling of 
the HMC algorithm, since the actual run was taking far too long. In table \ref{speedup} we  
see the speed-up of the different computing modes over a single CPU thread.
The CPU runs were performed using the Intel Fortran Compiler with the intel MKL while for the 
GPU runs we used the Nvidia CUDA compiler (nvcc). For these runs we found that the performance of the C2050 card is 
roughly equivalent to the performance of 96 threads of the cluster (for lattice size 24$^4$).

\begin{figure}
\includegraphics[width=0.7\columnwidth, angle=-90]{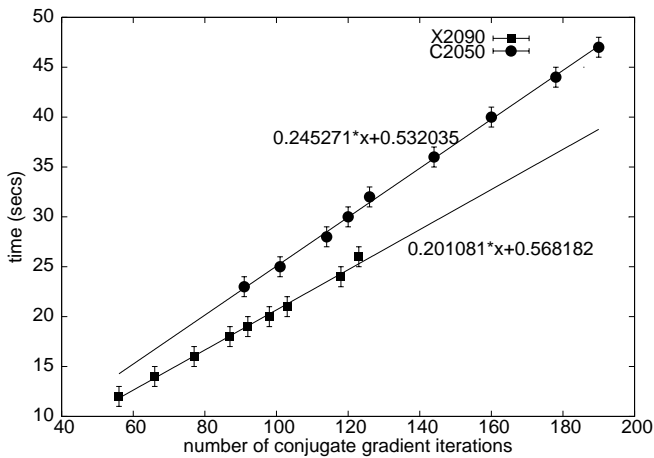}
\caption{Timing for the conjugate gradient routine vs number of conjugate gradient iterations. Error
bars represent the resolution of the timing measurements (1 sec).  The
intercept of the graph gives an estimate of the time taken to launch the kernel.}\label{gpu-scale}
\end{figure}

\begin{figure}
\includegraphics[width=0.7\columnwidth, angle=-90]{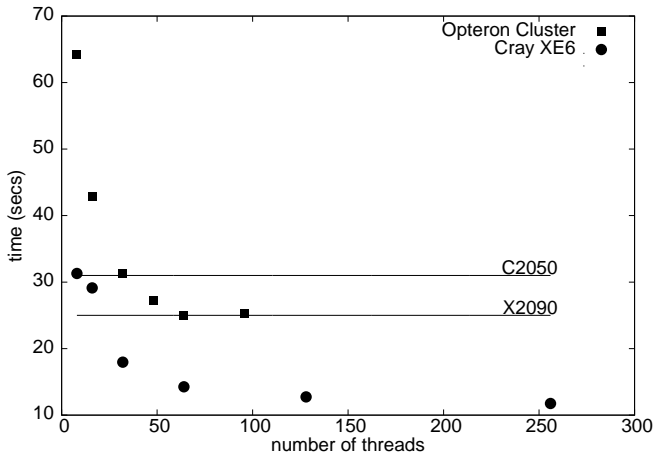}
\caption{Timing comparison between the CPU and the GPU inverter. Timings are for a CG inversion requiring 
123 iterations. Two different CPU runs viz. Opteron cluster with Infiniband interconnect and Intel Fortran 
Compiler and Cray XE6 with Cray Fortran Compiler, are shown along with the timings for a C2050 and a X2090 
GPU.}\label{inv_comp}
\end{figure}

We now report the timing information we have on the Cray hardware. Each Cray XE6 node consisted of 
two 16 core Opteron CPU at 2.1 GHz. On this 
machine the CPU code was compiled by the Cray fortran compiler and Cray libraries were used. 
The run parameters on the Cray were the same as the Opteron cluster. However instead of 200, only 2
trajectories were run starting from an ordered start.

We first tested the scaling of this program with the Intel Fortan Compiler on the XE6 and found that 
the performance was roughly the same as the Opteron cluster upto 64 threads. Beyond 64 threads the 
performance gain continued on the XE6, but not on the Opteron. 

The GPU run on the Cray was done for a different CG convergence limit 
compared to the CPU runs ($10^{-5}$ instead of $10^{-8}$) and so we could not directly compare the timing 
of the X2090 run with the other runs. Nevertheless, since the CG inversion timing is linear with 
the number of CG iterations for both GPUs (see figure \ref{gpu-scale}), it was possible 
to obtain a rough estimate for the timing for the X2090 for the same number of CG iterations as the other 
runs. An additional information that we get from figure \ref{gpu-scale}
is that the overhead for the CG routine is about 0.5 secs for both the C2050 and X2090.

In figure \ref{inv_comp} we compare the inverter timings for the two GPUs as well as the Opteron cluster and 
the Cray XE6 for a fixed number (123) of CG iterations. The C2050 gives the performance of 8 Cray threads or 
32 cluster threads while the X2090 gives a performance of about 20 Cray threads or 64 cluster threads.
 
\begin{figure}
\includegraphics[width=0.7\columnwidth, angle=-90]{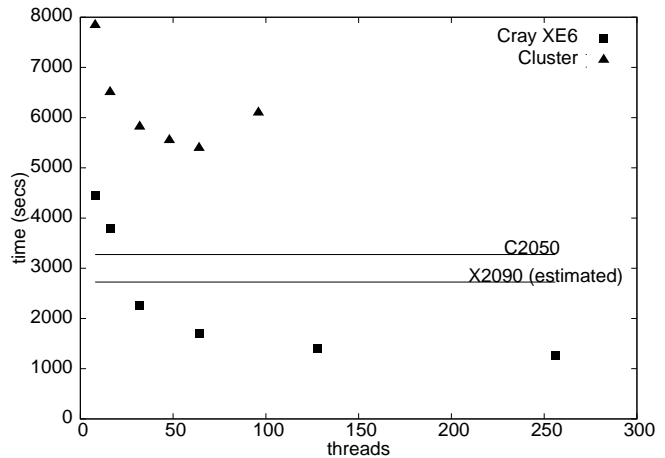}
\caption{Timing comparison between the Cray XE6 running Cray fortran compiler, opteron cluster with Infiniband 
interconnect running Intel Fortran compiler and the GPU cards C2050 and X2090 running nvcc and CUDA.}\label{time_comp}
\end{figure}

Figure \ref{time_comp} shows the full timing of the runs for the cluster, the XE6, the C2050 and the X2090. 
From this plot we conclude that the performance of the C2050 is roughly equivalent to about 19 
threads of the XE6, while that of the X2090 is that of about 26 threads.   

On the Cray XE6 nodes, the force calculation time was about 30\% of the inversion time with 64 threads, 
with 128 threads about 60\% of the inversion time and the 256 thread inverter took 
60\% less time than the force calculation. Therefore, on hindsight, on a lattice of size $24^4$ it is worth 
unrolling the loops over the force calculations as with the inverter beyond 64 threads.
On the Opteron cluster on the other hand the force calculation time was never more than 7\% of the inversion time. 
Thus there not much is to be gained by unrolling the force loops. 

Our conclusion is that GPU programs can be speeded up to a great extent by using the register variables 
instead of the global memory. To effectively use the register variables we had to map array elements to 
scalar variables. This was carried out automatically by a code generator. Before doing this mapping the 
performance gain was a modest factor of two. Using the registers effectively pushed up the performance 
gain to a factor 10 compared to a single thread performance on the lattice volumes we investigated. Such 
automated techniques might prove useful in the design of accelerator compilers. 

Here we did not use optimization techniques like mixed precision solvers or reconstructing the gauge link 
on the fly. We believe these optimizations can bring even more gains. 

At the moment we have tested the performance only on a single GPU. We hope to report our results on multiple 
GPUs shortly. 

\section{Acknowledgments}
The authors would like to acknowledge the DST (India) grant no. SR/S2/HEP/0035/2008 for funding the servers and GPU 
cards on which the programs were run. The authors would also like to acknowledge Cray Supercomputers for testing 
the program on their hardware and sharing the results with the authors.

\end{document}